# PEEPSS: Photonic-Enabled ExoPlanet Spectroscopic Sensor for the Habitable Worlds Observatory

Genevieve Markees[*,a,b], Stephen S. Eikenberry[a,b], Rodrigo Amezcua-Correa[a], Miguel Bandres[a], Rebecca Jensen-Clem[c], Sergio Leon-Saval[d], Laurent Pueyo[e], Raphaël Pourcelot[f]

[a]College of Optics and Photonics (CREOL), University of Central Florida, Orlando, FL, 32816,;
[b]Department of Physics, University of Central Florida, Orlando, FL, 32816;
[c]University of California Santa Cruz, Santa Cruz, 95064;
[d]School of Physics, University of Sydney, Sydney, Australia;
[e]Space Telescope Science Institute, Maryland, 21218;
[f]Max-Planck-Institut für Astronomie, Heidelberg, Germany

## ABSTRACT

The next few years will be critical for technology development for Habitable Worlds Observatory (HWO) in its mission to search for and characterize extrasolar planets. To achieve its stated goals with contrasts of one part in ten billion, HWO will require outstanding stability and precision, particularly in measuring and controlling the wavefront of the light propagate through the telescope and coronagraph system. We present simulations for the Photonic-Enabled ExoPlanet Spectroscopic Sensor (PEEPSS), which uses a set of photonic lanterns to efficiently couple light from the "dark hole" in the coronograph focal plane (where the exoplanets are expected to lie) into single-mode fibers and the main spectrograph. PEEPSS uses rejected host star light from the region interior to the dark hole to aid in the wavefront sensing; this has the advantage of doing the sensing in the coronograph focal plane, eliminating non-common-path errors between the wavefront sensing and science channels. The photonics lanterns allow us to combine our science channel and wavefront sensor into a single system. PEEPSS will be particularly advantageous provide in the near-infrared (NIR) bandpass, which is of particular interest for HWO. Because the limiting inner working angle (IWA) of a coronagraph scales as wavelength over diameter, exoplanet imaging in the NIR becomes a major challenge as the IWA can exceed the exoplanet orbital radius. PEEPSS will enable NIR coronagraphic observations at smaller IWA than other approaches, increasing the observational parameter space HWO can probe in the search for exoplanets.

## 1. INTRODUCTION

Over the past decade, the study of exoplanet atmospheres has expanded rapidly. Understanding the composition of these atmospheres, particularly around small, rocky exoplanets, is an important step in determining the potential habitability of these exoplanets. One way that this is presently done is through transmission spectroscopy, where the absorption spectrum of a transiting exoplanet's atmosphere is measured as it passes in front of its host star [1]. Although the study of smaller exoplanets' atmospheres is becoming possible from the ground, space-based telescopes, like James Webb Space Telescope, are particularly important in this endeavor, as the absorption from the Earth's (life-supporting) atmosphere does not interfere with the measurements of the exoplanet's atmosphere.

Another growing method of detecting and characterizing exoplanets is coronagraphy, where the light from the host star is suppressed, allowing dimmer companions to be directly imaged. Coronagraphs achieve this by redirecting the on-axis light (the star) while allowing the off-axis light (the planet) to pass through the optics to the focal plane. In general, this results in an annulus where the contrast between the star and exoplanet is greatest, sometimes called the "dark hole". However, even small misalignments can allow light from the star to leak into the dark hole, compromising the instrument's ability to detect the dim exoplanets [2]. Thus, controlling the wavefront of the light as it propagates through the system is critical, and remains a key focus in coronagraph design [3]. Wavefront control is often performed with deformable mirrors (DMs) which can correct the incoming wavefront. Spectroscopy of directly imaged exoplanets has some advantages over transmission spectroscopy, not only because of the need to disentangle the spectra of the star and exoplanet atmospheres, but also because different biosignatures are more detectable with one technique than with the other [1].

---

[*] genevieve.markees@ucf.edu

One upcoming mission that will have a significant impact on the study of exoplanet atmospheres and habitability is the proposed Habitable Worlds Observatory (HWO). A key goal for HWO will be to directly image and spectroscopically characterize ~25 Earth-like exoplanets, and in order to achieve this goal, it will require phenomenally stable starlight suppression. Designs for HWO's coronagraph are still under consideration, with special interest in parameters including inner working angle (IWA), throughput, and wavelength coverage [4]. Finding Earth-like exoplanets will be a challenge even for an instrument like HWO, because they are small, dim, and typically close to their host stars.

This paper describes the Photonics-Enabled ExoPlanet Spectrosopic Sensor (PEEPSS) for use on HWO, and presents some preliminary simulations of its wavefront sensing capabilities. PEEPSS uses a photonic fiber technology called the photonic lantern to address some of the challenges in high contrast coronagraphy. The stability and efficiency which photonic lanterns offer can help with the technological problems which remain a challenge for HWO. In addition, the system will also provide the capability to perform simultaneous wavefront sensing, imaging, and spectroscopy. Section 2 will briefly review the proposed design, which uses a series of photonic lanterns to split the incoming light. Building on that, section 3 will present simulations of the coupling of the post-coronagraph light into the photonic lanterns and the image field reconstruction, which can be used for both wavefront sensing and science imaging.

## 2. THE BASIC DESIGN OF PEEPSS

**A key component for PEEPSS – the photonic lantern**

The PEEPSS design relies on the photonic lantern, which is regularly highlighted as a key technology for the growing field of astrophotonics [5]. Originally developed with infrared, ground-based observations in mind, the photonic lantern acts as a waveguide, transforming one multimode image into a series of single-mode outputs; this allows photonic lanterns to combine the broadband coupling of a multimode fiber with the output stability and efficiency of single-mode fibers [6]. At visible wavelengths, photonic lanterns have broadband efficiencies >90%, making them well suited for spectroscopic work [7]. At present, numerous projects are underway to apply photonic lanterns to a variety of astronomical problems, such as adaptive optics wavefront sensing and spectro-astrometry of protoplanets [8-9].

PEEPSS uses both standard photonic lanterns, which couple well to linearly-polarized (LP) modes, and ring photonic lanterns, which are designed to couple to orbital angular momentum (OAM) modes. Examples of both these mode sets can be found in Figure 1. The ring lantern, which has a core with an annular cross-section rather than a circular one, was originally developed for communications purposes because OAM modes can be used for increased speed and security [10-11]. However, this shape is useful for the HWO and PEEPSS case because it is well-adapted for coupling to the dark hole of the coronagraph.

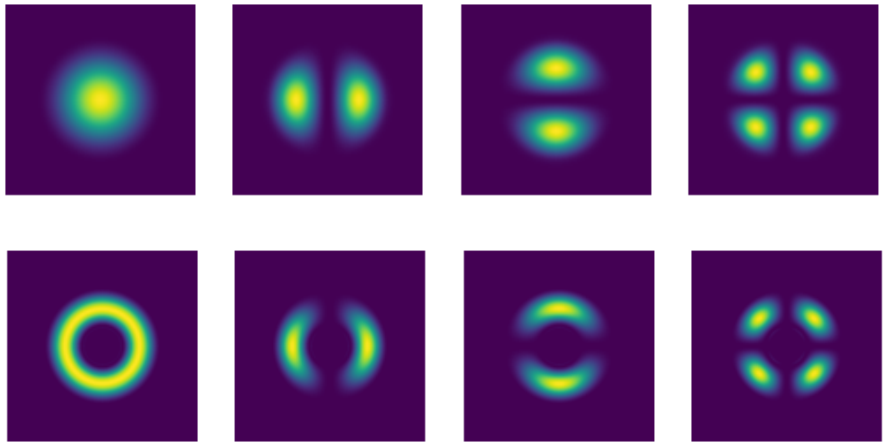

Figure 1. On the top, sample LP modes (the basis for the standard lantern), and on the bottom, sample OAM modes (the basis for the ring lantern).

Because photonic lanterns decompose a multi-mode light field into a set of discrete modes at high efficiency, reconstructing the complete incoming light field based on the output modes is possible. This includes retrieving the phase and amplitude of the light field, with the help of a Gerchberg-Saxton-style algorithm, rather than merely the intensity.

**The basic layout**

The PEEPSS module will be placed after the coronagraph pupil, and can act either as the primary science focal plane or as an alternative imaging plane. Figure 2 shows an example of PEEPSS for a single channel; proposed designs for HWO include multiple channels for the coronagraph, typically covering the near-infrared, visible, and ultraviolet [12-13]. PEEPSS is intended to work in concert with other wavefront sensors, providing a full phase-sensitive wavefront sensing capability at the "spectroscopic" focal plane.

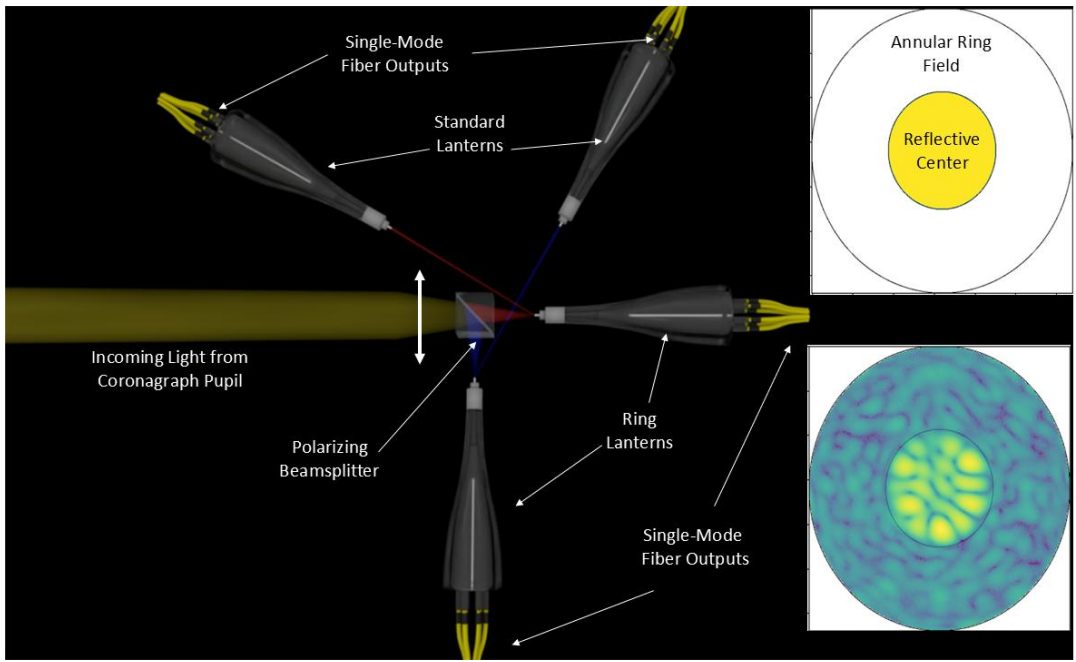

Figure 2. Layout for a PEEPSS module (single channel). Starting on the left, light from the coronagraph passes through focusing optics and a polarizing beamsplitter, before reaching the focus at the faces of the two ring lanterns. The dark hole couples to the ring lantern, while the central or "core" region reflects to another standard lantern. (Inset images) The top image shows a labelled representation of the face of the ring lanterns, while the bottom shows the same overlayed with a simulated coronagraph image.

First, the light from the coronagraph is split by a polarizing beamsplitter; this is necessary because photonic lanterns are polarization sensitive. PEEPSS splits the light across four photonic lanterns, two sets of a paired ring lantern and standard lantern (one set for each polarization). The ring lantern will redirect the center section of the coronagraph's light field to the standard lantern by means of a mirror coating on the center of its front face; this mirror coating will cover a region where light would not couple into the ring lantern. The light from the coronagraph's dark hole couples into the ring lantern, and the residual starlight in the center will be coupled into the standard lantern.

From there, each lantern decomposes the light field into the component spatial modes, and those modes pass through to a spectrograph, providing us with full spectral information covering the mode space of a given lantern. These spectra contain the full amplitude and phase information required to reconstruct the light field from the coronagraph. We assume an overall system throughput of ~15%, and noise-free detectors, as HWO anticipates using virtually noiseless photon-counting detectors [12,14].

Because the center field will be dominated by residual starlight, we plan to use this to aid in our wavefront sensing. The ring field, where the dark hole of the coronagraph is coupled, is where HWO will search for exoplanets. The placement of the PEEPSS module behind the coronagraph provides an important bonus, as this will provide the only fully common path wavefront sensing on HWO.

## 3. SIMULATIONS

Simulations of the various PEEPSS functions (wavefront sensing and spectroscopic imaging) begin with the same basic set up. Using the open-source HCIPy package https://docs.hcipy.org/0.5.1/, currently under development by the Exoplanet Exploration Program Coronagraph Design Survey team [15], we generate complex electric field distributions representing the light field at the coronagraph's focal plane. From there, we simulate the coupling of that electric field into the eigenmodes of a given type of lantern, using in-house modeling software which simulates the modal evolution of light field as it passes through the lantern. We apply a simulated transfer matrix (TxMx) to convert those eigenmodes into the output modes that we would measure from a lantern.

**Wavefront sensing performance**

First, we perform preliminary simulations based only on the ring portion of the lantern. Note that this simulation does not even make use of the residual starlight at the center of the field. We are able to reconstruct the true modal composition of the dark hole with high fidelity, as shown in the main plot of Figure 3; the blue line is the direct decomposition of a sample coronagraph dark hole light field, and the orange points are reconstructed using the TxMx method described above. This simulation assumes a V=1 mag reference star, with observations of 100 seconds and a 6-meter diameter aperture.

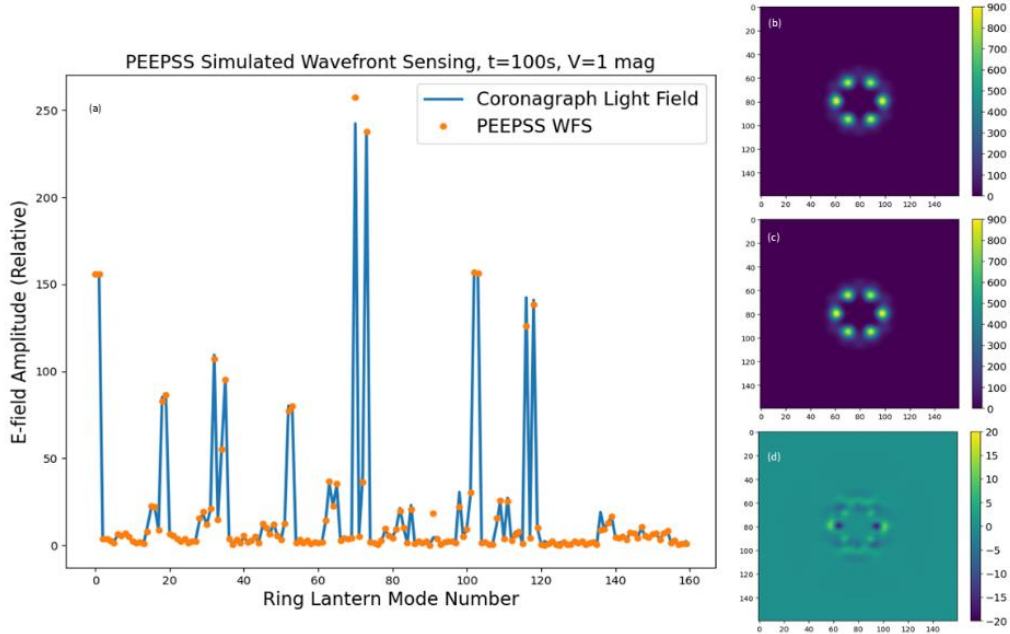

Figure 3. Sample wavefront reconstruction results, assuming a V=1 mag reference star observed for 100 seconds. (a) Modal decomposition of the dark hole light field in blue, overplotted with simulated PEEPSS reconstruction. (b) The original dark hole light field, (c) the reconstructed dark hole light field, and (d) the difference between (b) and (c), all in units of photons.

The reconstructed images for the V=1 mag reference star are dominated by shot noise, as are further simulations with V=6 mag science target stars. Because we can reach a high signal-to-noise ratio over relatively short (100 second) observations, PEEPSS should have the ability to rapidly detect and characterize aberrations in a wavefront. With a single 100 second image, PEEPSS provides wavefront information that can enable a factor of 10 improvement in contrast, as shown in the difference image Figure 3 (d).

Next, we simulate the complete proposed wavefront sensing approach for PEEPSS, using the residual starlight at the center of the coronagraph field. In this case, we use a series of 2300 electric field images with James Webb Space Telescope aberrations which were propagated through a simulated coronagraph [16]. For the first 2000 images, we use a matrix factorization technique called single value decomposition, which allows us to condense the information about the aberrations down into 50 "eigen field" images. These eigen fields can be used to create a core-annulus matrix which relates the center modes from the coronagraph field to the dark hole. Having in effect "trained" this core-annulus matrix on the first 2000 images from the set, we take the remaining 300, and apply the mapping matrix to the core modes of those. This process is presented in Figure 4, going around clockwise from the top left. This process reproduces the aberrations present in the dark hole with high fidelity, and would allow HWO to track and characterize wavefront aberrations simultaneously with science observations.

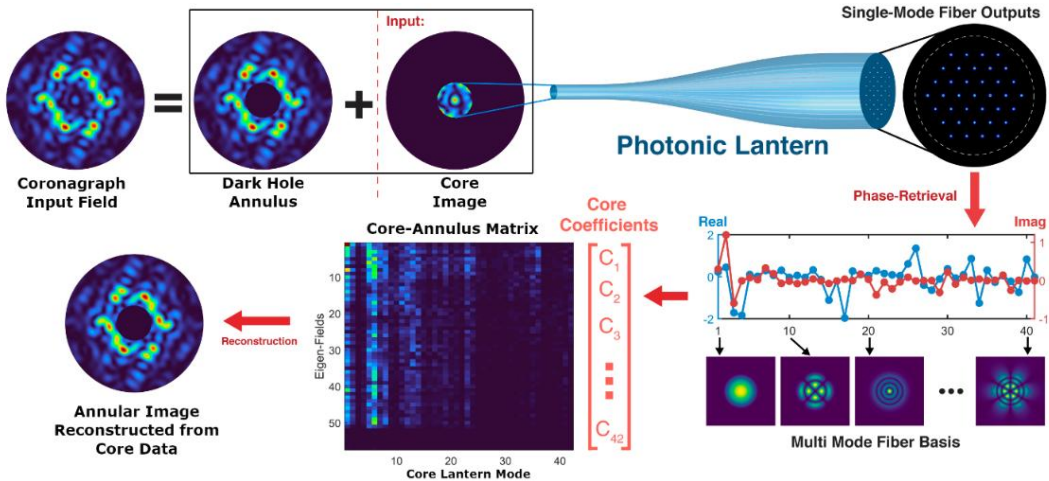

Figure 4. The complete proposed wavefront sensing process. (Starting from the top left and going around clockwise) The full coronagraph input field is separated into the dark hole and the core. A standard lantern splits the core region into spatial mode outputs, upon which we can perform a phase retrieval, giving us the phase and amplitude for each spatial mode. Using a previously calculated matrix to relate the core and the "eigenfields" (which describe the dark hole modes), we reconstruct the dark hole modes and light field based solely on the light in the core. Note that the "Dark Hole Annulus" and the "Annular Image Reconstructed from Core Data" are very similar, but the latter was reconstructed without any information contained in the original dark hole.

**Spectroscopic imaging performance**

PEEPSS can also be used to perform spectroscopic imaging, similar to an integral field unit (IFU). The key difference between an IFU and PEEPSS is that while an IFU maintains a separation between the light from different sources, PEEPSS will linearly combine light from each source in the lantern mode space. However, as shown in Figure 5, the overall signal-to-noise (SNR) of the PEEPSS approach should not differ meaningfully from the IFU approach in the case where measurements are limited by the target shot noise. In the detector noise limit, PEEPSS will experience slightly reduced SNR (on the order of a few tens of percent) compared to an IFU since the planet light is spread across a larger number of lantern modes (and thus pixels).

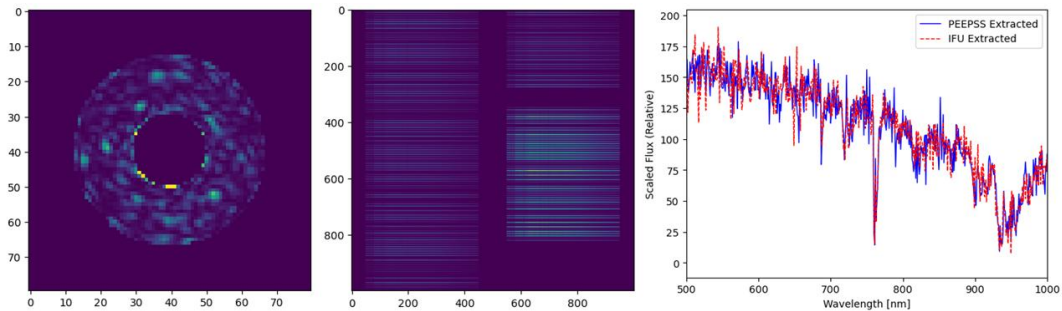

Figure 5. (Left) Dark hole with simulated exoplanet PSF which has a contrast of $3 \times 10^{-8}$, seen as the bright spot in the lower right quadrant. (Middle) Simulated PEEPSS spectrogram of the outputs of one ring lantern and one standard lantern, with one spectrum per mode. (Right) Simulated PEEPSS spectrum, overplotted with IFS spectrum for comparison. The noise

properties of these two approaches are similar. Note that the intensities shown here were numerical choices for the purpose of realistic simulation and should not be taken as true astrophysical values.

Mathematically, the SNR will not differ in the case of shot noise limited observations because both PEEPSS and an IFU will experience weighted shot noise from the star. When dealing with an IFU, we would apply a weight to each pixel based on the expected point spread function (PSF) of the exoplanet; then we would remove the average background induced by the star but are left with the PSF-weighted shot noise from that background starlight. For PEEPSS, we apply a weight to each pixel based on the modal distribution function (MDF), the expected distribution of the exoplanet light over the set of lantern modes. Just like in the IFU case, we subtract the average background starlight and are left with the MDF-weighted shot noise from the background starlight. Although the weights are not identical, spectra reconstructed from PEEPSS should have similar noise properties to comparable spectra collected with an IFU.

## 4. CONCLUSIONS

PEEPSS provides many benefits for HWO, both as a wavefront sensor and in combination as a science channel. Using the phase and amplitude information that photonic lanterns provide will allow PEEPSS to rapidly detect and characterize aberrations while digging the dark hole. In addition, the ability to continue to characterize wavefront aberrations during science operations will be a significant advantage for HWO's science productivity. Because of the phase and amplitude reconstruction – made possible by the photonic lanterns – PEEPSS can also provide simultaneous imaging and spectroscopy of exoplanets, once again cutting down on time requirements for observations of a given target. The lightweight design of PEEPSS combined with the ability to perform wavefront sensing and science observations simultaneously makes it a potentially powerful tool for HWO

Further work on the PEEPSS design is ongoing. Refining the lantern design to maximize our reconstruction fidelity is a key task; the precise number of modes for the standard and ring lanterns can impact how well the light fields couple, so we want to fully optimize that. In addition, other properties of the lanterns like the numerical aperture can also have an impact on coupling efficiency, and will be examined to ensure we are achieving optimum coupling.

## ACKNOWLEDGEMENTS

This material is based upon work supported by the AFRL Grand Challenge in Quantum-Inspired Imaging and the Air Force Office of Scientific Research under award number FA9550-24-1-0332.